\documentclass[aps,pra,twocolumn]{revtex4}
\usepackage{graphicx}
\usepackage{amsmath}

\begin{document}

\title{Transient first-order interference of two independent thermal light beams}

\author{Jianbin Liu}
\email[]{liujianbin@mail.xjtu.edu.cn}

\affiliation{Electronic Materials Research Laboratory, Key
Laboratory of the Ministry of Education \& International Center for
Dielectric Research, Xi'an Jiaotong University, Xi'an 710049, China}

\begin{abstract}
By analyzing the first-order interference of two independent thermal light beams with both classical and quantum theories, we conclude that it is impossible to observe the transient first-order interference pattern by superposing two independent thermal light beams even if the degeneracy parameter of thermal light is much greater than one. The result suggests that the classical model of thermal light field within the coherence time may not be the same as the one of laser light field within the coherence time.
\end{abstract}

\pacs{42.25.Hz, 42.25.Bs}

\date{\today}

\maketitle

Shortly after the invention of laser \cite{laser}, the first-order interference of two independent laser light beams was reported \cite{javan,magyar-nature,lipsett}. Magyar and Mandel observed spatial transient first-order interference pattern by superposing two independent ruby laser light beams \cite{magyar-nature}. Transient first-order interference pattern is the first-order interference pattern obtained in a short time interval, which is usually shorter than the coherence time of the field. Further more, Pfleegor and Mandel proved that the interference of two independent laser light beams takes place even under conditions in which ``the intensities are so low that, with high probability, one photon is absorbed before the next one is emitted by one or the other source'' \cite{pfleegor}. Forrester \textit{et al.} observed beats by mixing Zeeman components of a visible spectral line \cite{forrester}. However, their experiment can not be regarded as the interference of two independent thermal light beams. For the interfering fields in their experiment have common origin, which is similar as the latter experiments of interference of light emitted by two sources \cite{cohen-1966,grangier,kuo,eichmann,afzelius}. The transient first-order interference of two independent thermal light beams like the one with two independent laser light beams \cite{javan,magyar-nature,lipsett} has never been reported. Most physicists attribute it to that the degeneracy parameter of thermal light is usually much less than one \cite{mandel-rmp,hecht}. On the other hand, if the degeneracy parameter of thermal light is much greater than one, the transient first-order interference pattern of two independent thermal light beams can be observed. Is this prediction true? Our answer is no. In the following part, we will show that it is impossible to observe the transient first-order interference pattern by superposing two independent thermal light beams even if the degeneracy parameter of thermal light is much greater than one. Our results suggest that thermal and laser light fields are different within the coherence time.

Thermal light is usually obtained by passing blackbody radiation through linear filters, such as apertures, mirrors, lenses, polarizers, \textit{etc} \cite{mandel-book}. It is also sometimes called chaotic light. Gas discharge lamp is one of the typical thermal light sources, where the different excited atoms emit their radiation independently of one another. The total thermal light field equals the sum of all these randomly and independently emitted radiation fields. For example, let us follow Loudon's book to discuss the temporal fluctuation of polarized thermal light emitted by collision-broadening thermal source \cite{loudon}. The field emitted by the $j$th atom can be written as
\begin{equation}
E_j(t)=E_0\exp{[-i2\pi \nu t+i\varphi_j(t)]},
\end{equation}
where $E_0$ and $\nu$ are the amplitude and frequency of the emitted field, respectively. The phase $\varphi_j(t)$ remains constant during periods of free flight but it changes abruptly each time a collision occurs \cite{loudon}. The total field produced by a large number of radiating atoms is
\begin{eqnarray}\label{thermal-field}
E(t)&=&\sum_{j=1}^{N}E_0\exp{[-i2\pi \nu t+i\varphi_j(t)]}\nonumber\\
&=&E_0\exp{[-i2\pi \nu t]}a(t)\exp{[i\varphi(t)]},
\end{eqnarray}
where the amplitude and frequency of the field are assumed to be the same for different atoms and different periods. $N$ is the number of radiating atoms. The amplitude $a(t)$ and phase $\varphi(t)$ are the results of statistical sum of a random walk process, which are \textit{different at different instants of time} \cite{loudon}. Hence the phase of the thermal light field emitted by collision-broadening thermal source is not a constant during the coherence time.

The total thermal light field can be treated as an incoherent superposition of Fourier modes \cite{shih-book}. For quasi-monochromatic thermal light emitted by collision-broadening thermal source, the field can be written as
\begin{eqnarray}\label{quasimono}
E(t)&=& \int_{-\infty}^{\infty}f(\nu)\exp{[-i2\pi \nu t+i\varphi_{\nu}(t)]}d\nu\nonumber\\
&=&A(t)\exp{[-i2\pi \nu_0t+i\Phi(t)]}.
\end{eqnarray}
Where $f(\nu)$ and $\nu_0$ are the spectrum and mean frequency of the field, respectively. $\varphi_{\nu}(t)$ is the phase of Fourier mode $\nu$ at time $t$. $A(t)$ and $\Phi(t)$ are the amplitude and phase of the quasi-monochromatic thermal light field at time $t$, respectively.

One should not confuse Eq. (\ref{quasimono}) with the expression for quasi-monochromatic light field given in Born and Wolf's book \cite{born}
\begin{eqnarray}\label{quasimono-born}
E(t)&=& \int_{-\infty}^{\infty}f(\nu)\exp{[-i2\pi \nu t+i\varphi(\nu)]}d\nu\nonumber\\
&=&A(t)\exp{[-i2\pi \nu_0t+i\Phi(t)]},
\end{eqnarray}
where we have changed the symbols in Born and Wolf's book so that the symbols in Eqs. (\ref{quasimono}) and (\ref{quasimono-born}) are consistent. The phase $\varphi(\nu)$ does not change with time $t$ in Eq. (\ref{quasimono-born}), while the phase $\varphi_{\nu}(t)$ may change with both the frequency $\nu$ and time $t$ in Eq. (\ref{quasimono}). Hence the conclusion that $A(t)$ and $\Phi(t)$ are substantially constant within the coherence time drawn in Born and Wolf's book may not be valid for quasi-monochromatic thermal light field.

In classical theory, we have seen that the phase of thermal light field emitted by collision-broadening thermal source is not a constant during the coherence time. If the phase of thermal light field changes rapidly over 0 and $2\pi$ during the measurement time, there will be no transient first-order interference pattern by superposing two independent thermal light beams. In quantum theory, we will show that this conclusion holds for thermal light emitted by more types of thermal sources. It is well-known that photons in thermal and laser light are emitted by spontaneous and stimulated emissions, respectively \cite{einstein}. It is reasonable to assume the initial phases of photons in thermal light are random and uniformly distributed between 0 and $2\pi$, while the initial phases of photons in a single-mode continuous wave laser light are identical within the coherence time \cite{mandel-book}.

Quantum theory has been employed to interpret the first-order interference of light for a long time \cite{bohm,feynman-lecture,mandel-1964}. In our earlier studies, we find it is helpful to understand the physics of the second-order interference light in Feynman's path integral theory \cite{liu-PRA,liu-OE}. We will also employ Feynman's path integral theory to interpret the transient first-order interference of two independent thermal light beams, hoping to understand the phenomenon better.

\begin{figure}[htb]
    \centering
    \includegraphics[width=70mm]{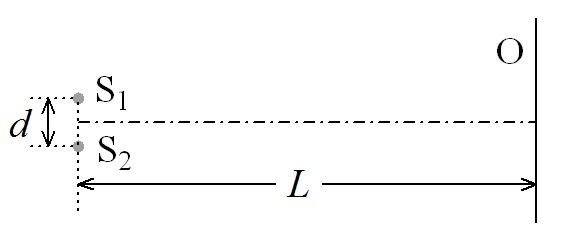}
    \caption{The scheme for the interference of two independent thermal light beams. $\text{S}_1$ and $\text{S}_2$ are two independent point thermal sources. $\text{O}$ is the observing plane. $d$ is the distance between $\text{S}_1$ and $\text{S}_2$. $L$ is the distance between the source and observing planes.
    }\label{young}
\end{figure}

The scheme for the interference of two independent thermal light beams is shown in Fig. \ref{young}. $\text{S}_1$ and $\text{S}_2$ are two identical but independent point sources emitting polarized quasi-monochromatic thermal light. For simplicity, we assume the light beams emitted by these two sources have equal intensities and consider one-dimension case only. There are two different ways for a photon to be detected at space-time coordinate $(x,t)$ on the observation plane. One is the detected photon is emitted by S$_1$ and the other one is the photon is emitted by S$_2$. In classical picture, one can always figure out the detected photon comes from S$_1$ or S$_2$, for the measurement accuracy can be arbitrarily high without influencing the system. However, it is not the case in quantum mechanics \cite{bohm}. Based on the conclusion that all the photons within the coherence volume are intrinsically indistinguishable from each other \cite{mandel-book}, it is straightforward to judge whether these two different ways are distinguishable or not. In the scheme shown in Fig. \ref{young}, the effective transverse coherence length of the field emitted by these two point sources can be treated as the same as the coherence length of the field emitted by a thermal source with dimension of $d$. If the uncertainty in the position detection is not greater than the transverse coherence length, $\lambda_0 L/d$, these two different ways are indistinguishable. Where $\lambda_0$ is the mean wavelength of the photon. In fact, the position uncertainty in the usual photon detection is much less than $\lambda_0 L/d$, which means these two different ways are usually indistinguishable if other properties of photons emitted by these two sources are identical.

When these two different ways to trigger a photon detection event on the observing plane are indistinguishable, the probability distribution for the $j$th detected photon is \cite{feynman-path}
\begin{equation}\label{pj}
P_j(x)=|e^{i \varphi_{j1}}K_1(x)+e^{i\varphi_{j2}}K_2(x)|^2.
\end{equation}
Where $K_\alpha(x)$ ($\alpha=1$, and 2) is the Feynman's photon propagator for the photon emitted by $\text{S}_\alpha$ goes to $x$ in the observing plane. $\varphi_{j1}$ and $\varphi_{j2}$ are the initial phases of the $j$th detected photon emitted by $\text{S}_1$ and $\text{S}_2$, respectively. The finally observed first-order interference pattern is proportional to the ensemble average of all these single photon probability distributions.

It is easy to find that the Feynman's photon propagator is the same as the Green function in classical optics \cite{feynman-path,born}. Hence the results are the same in both quantum and classical calculations. The only difference is the interpretation. We will directly employ the results in classical optics in the following discussions. Equation (\ref{pj}) can be simplified as \cite{born}
\begin{equation}\label{pj-simplify}
P_j(x)\propto 1+\cos(\frac{2\pi d}{\lambda_0 L}x+\varphi_{j1}-\varphi_{j2}),
\end{equation}
in which periodic modulation of the probability distribution is obvious. However, it is impossible to observe interference pattern with only one photon. One has to collected certain number of photons to observe interference pattern. Since the initial phases of photons in thermal light are random and uniformly distributed between 0 and $2\pi$. The relative phase, $\varphi_{j1}-\varphi_{j2}$, changes randomly for every detected photon. This conclusion is true even if all the photons are detected in a time interval shorter than the coherence time.

\begin{figure}[htb]
    \centering
    \includegraphics[width=70mm]{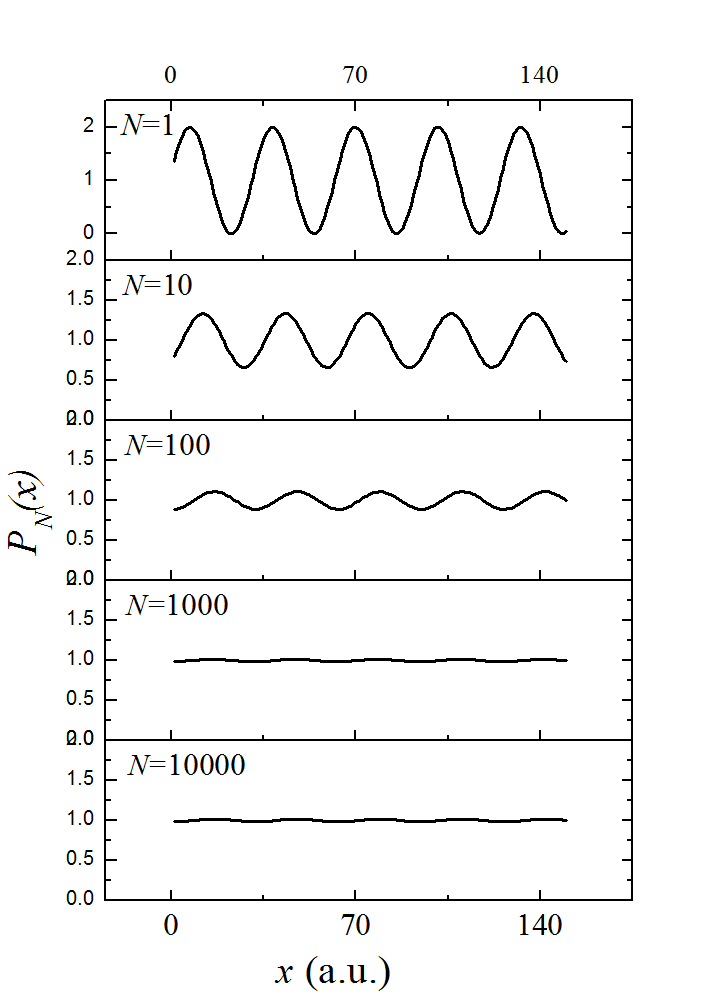}
    \caption{Photon probability distributions for different number of detected photons when superposing two independent thermal light beams. $N$ is the number of detected photons. $P_N(x)$ is the photon probability distribution for $N$ detected photons. The periods of all the probability distributions are the same. However, the positions of the maximums and minimums are random for different simulations.
    }\label{patterns}
\end{figure}

The probability distribution for a finite number of photons is given by the sum of $N$ different single-photon probability distributions,
\begin{equation}\label{pn}
P_N(x)\propto \sum_{j=1}^{N}[1+\cos(\frac{2\pi d}{\lambda_0 L}x+\varphi_{j1}-\varphi_{j2})].
\end{equation}
Since the relative phase, $\varphi_{j1}-\varphi_{j2}$, is random for every detected photon, it is also a random walk problem. With the same method in Refs \cite{hadzibabic,ashhab,cennini}, it is straightforward to get the probability distribution for different number of photons as
\begin{equation}\label{pn}
P_N(x)\propto 1+\frac{1}{\sqrt{N}}\cos(\frac{2\pi d}{\lambda_0 L}x+\varphi),
\end{equation}
where $\varphi$ is a random phase determined by the sum of all $N$ different relative phases. Figure \ref{patterns} shows the simulated probability distributions for different number of photons. There is periodic distribution for any finite number of photons. However, the visibility of the periodic probability distribution decreases rapidly as the number of photons increases. As $N$ goes to infinity, $P_N(x)$ becomes a constant, which indicates that the visibility  of the first-order interference pattern equals zero.

\begin{figure}[htb]
    \centering
    \includegraphics[width=70mm]{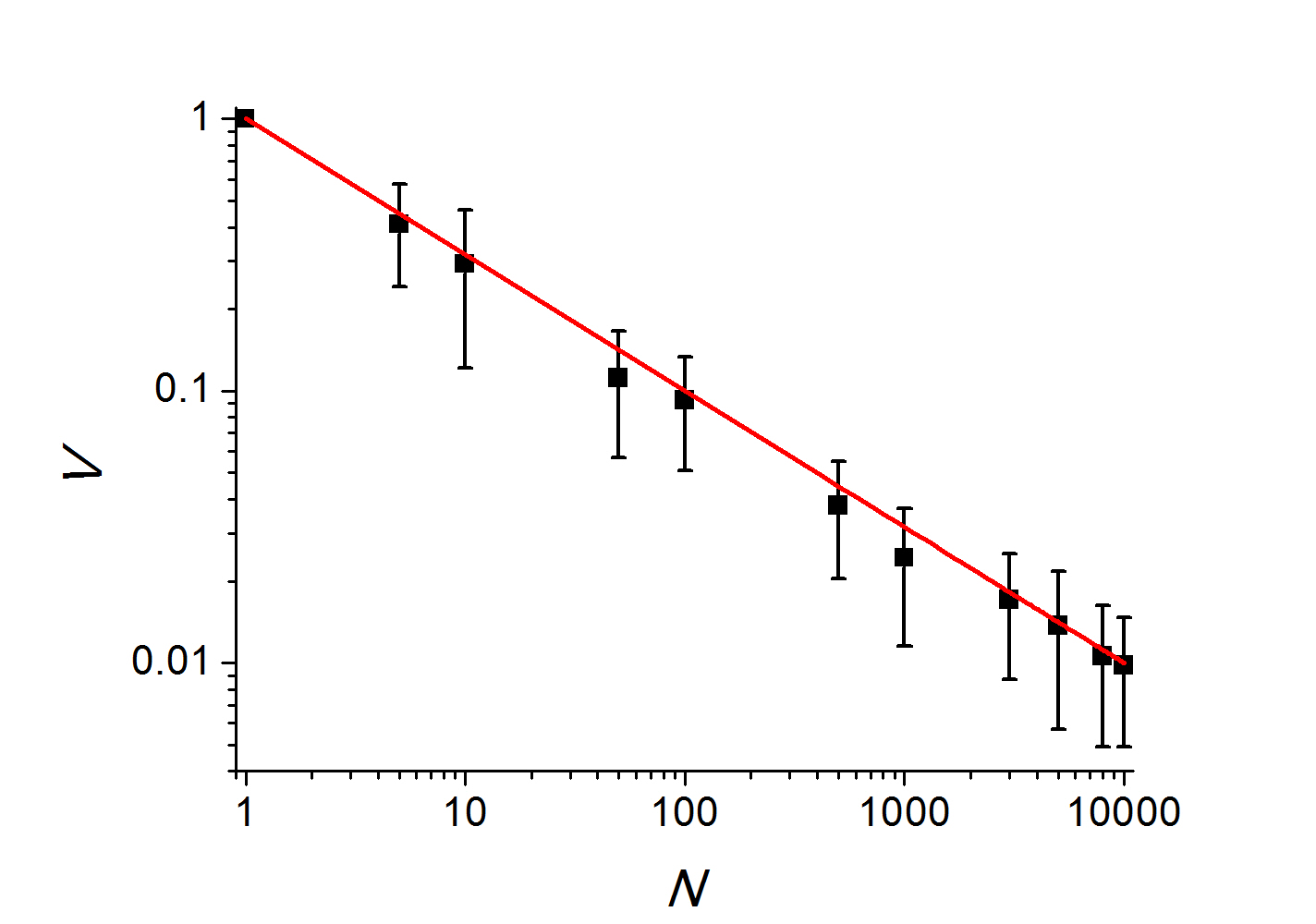}
    \caption{(Color online) Visibility of photon probability distribution for different number of photons when superposing two
    independent thermal light beams. $V$ is the visibility and $N$ is the number of detected photons. Each dot is the average of 50 independent numerical trials. The red solid line is the theoretical curve of $V=1/\sqrt{N}$.
    }\label{visibility}
\end{figure}

Based on Eq. (\ref{pn}), the visibility of the probability distribution for different number of photons is given by
\begin{equation}\label{visi-formula}
V=\frac{1}{\sqrt{N}}.
\end{equation}
Figure \ref{visibility} shows the theoretical simulated visibility for different number of photons, in which each dot is an average of 50 independent numerical trials and the red line is the theoretical curve of $V=1/\sqrt{N}$. The theoretical result is consistent with the numerical simulations within the errors. One should not confuse the conclusion here with the one of thermal light in a Young's double-slit interferometer, in which, the visibility of first-order interference pattern is independent of the number of detected photons.

When the number of detected photons is small, the visibility of the photon probability distribution is high. However, there may not be enough photons to retrieve the probability distribution. On the other hand, when the number of photons is large enough to retrieve the probability distribution, the visibility may be too low to show the interference pattern. This is similar as the first-order interference pattern of two independent laser light beams disappears for long measurement time. Hence we may conclude that, in the model that the initial phases of all photons in thermal light are random and uniformly distributed between 0 and $2\pi$, it is impossible to observe the first-order interference pattern by superposing two independent thermal light beams. The conclusion is true for both short and long measurement time compared to the coherence time of the thermal light.

Comparing classical and quantum interpretations for the transient first-order interference of two independent thermal light beams above, the conclusion that the phase $\Phi(t)$ of thermal light field is not a constant during the coherence time should be valid for thermal light emitted by all types of thermal sources in which the initial phases of the photons are random. One would argue that if $\Phi(t)$ of thermal light field is not a constant during the coherence time, how can we observe interference pattern of thermal light in the usual Young's double-slit interferometer or Michelson interferometer? It is true that this classical model can not interpret the phenomenon. However, the existed interference pattern in these two interferometers is a result of the field emitted by a atom interferes with itself \cite{born}. $\Phi(t)$  is not necessary to be a constant. There is indeed one way to testify whether  $\Phi(t)$ of thermal light field during the coherence time is constant or not, which is the transient interference of two independent thermal light beams. If there is transient first-order interference pattern by superposing two independent thermal light beams, just like the one with two independent laser light beams \cite{magyar-nature}, $\Phi(t)$ has to be a constant during the coherence time. On the contrary, if there is no transient interference pattern, $\Phi(t)$ is not a constant. As we have shown above, there is no transient first-order interference pattern by superposing two independent thermal light beams. Hence $\Phi(t)$ of thermal light field is not a constant during the coherence time.

The difference between the transient first-order interference by superposing two independent thermal and laser light beams is more obvious if we analyze both of them in Feynman's path integral theory. The initial phases of photons in a single-mode continuous wave laser light are identical within the coherence time \cite{einstein}. The probability distribution for the $j$th detected photon by superposing two independent laser light beams is also given by Eq. (\ref{pj-simplify}). Unlike in the thermal light case, the relative phase in the laser light case will not change for different detected photons during the coherence time. The probability distribution function for a finite number of photons is the same as single-photon probability distribution, Eq. (\ref{pj-simplify}). Hence there will be transient interference pattern by superposing two independent laser light beams \cite{magyar-nature}.

Due to the degeneracy parameter of thermal light is much less than one, it is impossible to receive enough photons within the coherence time to experimentally testify our conclusion. Pseudothermal light can not be employed to test the conclusion either, for the initial phases of photons are not random during the coherence time for pseudothermal light \cite{martienssen}. However, there is an alternative way to testify the conclusion, which is by employing the cold atoms just above the threshold temperature of BEC \cite{leggett,cornell}. It has been proved that there is first-order interference pattern by superposing two independent BECs \cite{andrews}, which is just the same as the transient interference pattern by superposing two independent laser light beams \cite{magyar-nature}. If there is a way to superpose two independent cold atomic beams within the coherence time, one could judge whether there is interference pattern or not, which is an analogy of superposing two independent thermal light beams.

In conclusion, we have proved that there is no transient first-order interference pattern by superposing two independent thermal light beams. The reason is not the degeneracy parameter of thermal light is much less than one, but the initial phases of the photons in thermal light are random. The transient first-order interference pattern by superposing two independent thermal light beams can not be observed even if there is large number of photons within the coherence volume. Our results suggest that the classical models for thermal and laser light field within the coherence time are different. Extreme care is necessary when employing classical model of thermal light field to interpret the first-order interference in the Young's double-slit or Michelson interferometer, ``or else one runs the risk of being seduced by the law of analogy'' \cite{sudarshan}.

\section*{Acknowledgement}
This project is supported by Doctoral Fund of Ministry of Education of China under Grant No. 20130201120013 and the Fundamental Research Funds for the Central Universities.

\end{document}